\documentclass[twocolumn,           
               showpacs,            
               preprintnumbers,     
               aps,                 
               prd,          	    
               a4paper,             
               superscriptaddress,  
               nofootinbib,         
               tightenlines,        
               floats               
               ]{revtex4}

\usepackage{graphicx}
\usepackage{latexsym}
\usepackage{amsmath,amssymb}        
\usepackage[draft=false]{hyperref}

\newcommand{\de}{\mathrm{d}}
\renewcommand{\(}{\left(}
\renewcommand{\)}{\right)}
\renewcommand{\[}{\left[}
\renewcommand{\]}{\right]}
\newcommand{\period}{\,\mathrm{.}}
\newcommand{\comma}{\,\mathrm{,}}
\newcommand{\reffig}[1]{Fig.~\ref{#1}}
\newcommand{\refeq}[1]{Eq.~(\ref{#1})}

\newcommand{\reftab}[1]{Table~\ref{#1}}

\newcommand{\mpl}{m_\mathrm{Pl}}

\newcommand{\abs}[1]{\left\vert#1\right\vert}

\renewcommand{\Re}{\mathrm{Re}}

\newcommand{\wf}{w_\mathrm{f}}
\newcommand{\wph}{w_\varphi}
\newcommand{\gf}{\gamma_\mathrm{f}}
\newcommand{\gph}{\gamma_\varphi}
\newcommand{\of}{\Omega_\mathrm{f}}
\newcommand{\oph}{\Omega_\varphi}
\newcommand{\rhf}{\rho_\mathrm{f}}
\newcommand{\rhph}{\rho_\varphi}
\newcommand{\rhtot}{\rho_\mathrm{tot}}
\newcommand{\pf}{p_\mathrm{f}}
\newcommand{\pph}{p_\varphi}
\newcommand{\ptot}{p_\mathrm{tot}}
\newcommand{\csph}{c_{\mathrm{s}\varphi}^2}
\newcommand{\cs}{c_\mathrm{s}^2}

\newcommand{\drf}{\delta\rho_\mathrm{f}}
\newcommand{\drph}{\delta\rho_\varphi}
\newcommand{\drtot}{\delta\rho_\mathrm{tot}}
\newcommand{\dpf}{\delta p_\mathrm{f}}
\newcommand{\dpph}{\delta p_\varphi}
\newcommand{\dptot}{\delta p_\mathrm{tot}}
\newcommand{\dpnad}{\delta p_\mathrm{nad}}
\newcommand{\dph}{\delta\varphi}
\newcommand{\dqf}{\delta q_\mathrm{f}}
\newcommand{\dqph}{\delta q_\varphi}
\newcommand{\dqtot}{\delta q_\mathrm{tot}}
\newcommand{\Vf}{\mathcal{V}_\mathrm{f}}
\newcommand{\delf}{\delta_\mathrm{f}}
\newcommand{\delph}{\delta_\varphi}
\newcommand{\delpi}{\delta_\pi}
\newcommand{\R}{\mathcal{R}}

\newcommand{\vv}{\mathbf{v}}
\newcommand{\MM}{\mathcal{M}}
\newcommand{\rhot}{\rho_\mathrm{tr}}
\newcommand{\gtp}{\gamma^{\scriptscriptstyle +}_\mathrm{tr}}
\newcommand{\Nm}{N_\mathrm{-}}
\newcommand{\Ni}{N_\mathrm{i}}
\newcommand{\Nk}{N_\mathrm{k}}
\newcommand{\NpI}{N_\mathrm{pI}}
\newcommand{\NpII}{N_\mathrm{pII}}
\newcommand{\Np}{N_\mathrm{+}}
\newcommand{\DNk}{\Delta\Nk}
\newcommand{\DNpI}{\Delta\NpI}
\newcommand{\DNpII}{\Delta\NpII}
\newcommand{\DNp}{\Delta N_\mathrm{p}}
\newcommand{\Vp}{V_\mathrm{+}}

\begin{document}


\title{Perturbations in cosmologies with a scalar field and a perfect fluid}
\author{Nicola Bartolo}
\affiliation{Astronomy Centre, University of Sussex, 
             Brighton BN1 9QH, United Kingdom}
\author{Pier-Stefano Corasaniti}
\affiliation{ISCAP, Columbia University, Mailcode 5247, 
             New York NY 10027, United States}
\author{Andrew R.~Liddle}
\affiliation{Astronomy Centre, University of Sussex, 
             Brighton BN1 9QH, United Kingdom}
\author{Micha\"el Malquarti}
\affiliation{Astronomy Centre, University of Sussex, 
             Brighton BN1 9QH, United Kingdom}
\date{\today} 
\pacs{98.80.Cq \hfill astro-ph/0311503}
\preprint{astro-ph/0311503}


\begin{abstract}
We study the properties of cosmological density perturbations in a 
multi-component system consisting of a scalar field and a perfect fluid. We 
discuss the number of degrees of freedom completely describing the system, 
introduce a full set of dynamical gauge-invariant equations in terms of the 
curvature and entropy perturbations, and display an efficient formulation of 
these equations as a first-order system linked by a fairly sparse matrix. Our 
formalism includes spatial gradients, extending previous formulations restricted 
to the large-scale limit, and fully accounts for the evolution of an 
isocurvature mode intrinsic to the scalar field. We then address the issue of 
the adiabatic condition, in particular demonstrating its preservation on large 
scales. Finally, we apply our formalism to the quintessence scenario and clearly 
underline the importance of initial conditions when considering late-time 
perturbations. In particular, we show that entropy perturbations can still be 
present when the quintessence field energy density becomes non-negligible.
\end{abstract}


\maketitle

\section{Introduction}

The material content of the universe is commonly assumed to be a mixture of 
fluids, such as radiation or non-relativistic matter, and scalar fields, either 
driving a period of early universe inflation \cite{infbook} or playing the role 
of dark energy (quintessence) in the present universe \cite{quint,RP,CDS}. The 
latter possibility has motivated a number of works devoted to the study of 
cosmological perturbations in a multi-component system consisting of fluids and 
a scalar field, for instance 
Refs.~\cite{RP,CDS,quintpert,nonminquipert,PB,BMR,AF,SA,ML2,DMSW}. Nevertheless, 
the literature contains some contradictory statements concerning the properties 
of such perturbations.

In this paper we aim to resolve these discrepancies and will provide a 
comprehensive analysis of the problem. We will study the role of intrinsic 
entropy perturbation in the scalar field, and whether the notion of adiabaticity 
is preserved by the dynamics of the multi-component system when the evolution of 
such an intrinsic entropy perturbation is explicitly accounted for. In the 
process of doing so, we will discuss the number of degrees of freedom which 
completely describe the system and we will find a highly-efficient formulation 
of the perturbation equations including the effects of spatial gradients. 

Finally, using our formalism we will specifically discuss the quintessence 
scenario. We will correct some common misconceptions, discuss the evolution of 
entropy perturbations, and clearly show the importance of initial conditions 
when 
considering late-time perturbations. In particular, we will show that entropy 
perturbations can be enhanced by the evolution of the field and may still be 
present when its density is no longer negligible. This is an important result 
which has 
generally been overlooked when studying dark energy models.

\section{The dynamical equations}

\subsection{Background}

Our approach builds on an earlier paper by Malquarti and Liddle \cite{ML2}, and 
we will largely follow the notation of that article but with some differences in 
definitions. We assume a flat Friedman--Robertson--Walker universe throughout, 
with the background evolution determined by the usual equations
\begin{eqnarray}
3H^2 \,=\, 3\Big(\frac{\dot{a}}{a}\Big)^2
  &=& 8\pi G\rhtot \comma\\
2\dot{H}+3H^2 \,=\, 2\frac{\ddot{a}}{a}+\Big(\frac{\dot{a}}{a}\Big)^2
  &=& - 8\pi G \, \ptot \comma
\end{eqnarray}
where $H\equiv\dot{a}/a$ is the Hubble parameter, $a$ is the scale factor, a dot 
stands for a derivative with respect to cosmic time $t$ and the subscript 
``tot'' always refers to the sum over all matter components. The fundamental 
ingredients we consider here are a perfect fluid with constant equation of state 
$\wf\equiv\pf/\rhf$ and a minimally coupled scalar field $\varphi$ with 
potential $V(\varphi)$. Since we treat the fluid and the scalar field as 
uncoupled, the conservation of their respective energy--momentum tensors gives
\begin{eqnarray}
\dot{\rhf}
  &=& -3H(1+\wf)\rhf \qquad\Rightarrow\quad \rhf\propto a^{-3(1+\wf)} \comma 
\label{rhfeq}\\
\ddot{\varphi}
  &=& -3H\dot{\varphi}-\frac{\de V}{\de \varphi} \period \label{varphieq}
\end{eqnarray}
The subscripts ``f'' and ``$\varphi$'' will always refer to the perfect fluid 
and the scalar field respectively. 

Useful parameters describing completely the background properties of the scalar 
field are its equation of state 
\begin{equation}
\wph \equiv \frac{p_\varphi}{\rho_\varphi}
 = \frac{\dot{\varphi}^2/2-V(\varphi)}{\dot{\varphi}^2/2+V(\varphi)} 
\comma
\end{equation}
and its adiabatic sound speed
\begin{equation}
\csph \equiv \frac{\dot{p}_\varphi}{\dot{\rho}_\varphi}
 = \wph-\frac{\dot{w}_\varphi}{3H(1+\wph)}
 = 1+\frac{2}{3}\frac{\de V/\de\varphi}{H\dot{\varphi}}
\period
\end{equation} 
We also introduce the total equation of state $w\equiv\ptot/\rhtot$ and the 
total sound speed $\cs\equiv\dot{p}_\mathrm{tot}/\dot{\rho}_\mathrm{tot}$, and 
in order to simplify some expressions we define for each component 
$\gamma_\mathrm{x}\equiv1+w_\mathrm{x}$ and also $\gamma\equiv1+w$.

\subsection{Perturbations}

We consider only scalar perturbations and we choose to work in Newtonian 
gauge \cite{MFB}, where the perturbed metric reads as
\begin{equation}
\de s^2=-(1+2\Phi)\de t^2+a^2(t)(1-2\Psi)\de \mathbf{x}^2\,.
\end{equation}
Here $\Phi$ and $\Psi$ describe the metric perturbation, and in this case are 
equal to the gauge-invariant potentials defined in Ref.~\cite{MFB}. We work in 
Fourier space and compute the first-order perturbed Einstein equations. As our 
system has no anisotropic stress, the $(i-j)$ Einstein equations imply that the 
metric potentials are equal, $\Psi = \Phi$. The remaining Einstein equations are
\begin{eqnarray}
-3H(H\Phi+\dot{\Phi})-\frac{k^2}{a^2}\Phi &=& 4\pi G\drtot \comma 
\label{e:ein1}\\
\ddot{\Phi}+4H\dot{\Phi}+(2\dot{H}+3H^2)\Phi &=& 4\pi G\dptot \comma
\label{e:ein2}\\
-(H\Phi+\dot{\Phi}) &=& 4\pi G\dqtot \comma
\label{e:ein3}
\end{eqnarray} 
where $\nabla\delta q_{\rm tot}$ is the total momentum perturbation of the 
system. Eq.~(\ref{e:ein1}) comes from the $(0-0)$ Einstein equation, 
Eq.~(\ref{e:ein2}) from the $(i-i)$ Einstein equation, while Eq.~(\ref{e:ein3}) 
is obtained from the $(0-i)$ Einstein equation. The perfect fluid and scalar 
field perturbation variables are
\begin{eqnarray}
\dpf  &=& \wf\drf \comma\\
\dqf  &=& \rhf\gf\Vf \comma\\
\drph &=&
      \dot{\varphi}\delta\dot{\varphi}-\dot{\varphi}^2\Phi
      +\frac{\de V}{\de \varphi}\dph \comma\\
\dpph &=& 
      \dot{\varphi}\delta\dot{\varphi}-\dot{\varphi}^2\Phi
      -\frac{\de V}{\de \varphi}\dph \comma\\
\dqph &=& -\dot{\varphi} \, \delta\varphi \comma
\end{eqnarray}
where $\Vf$ is the fluid velocity potential defined so that the fluid velocity 
is given by $\delta\mathbf{u}\equiv\nabla\Vf$ --- this is possible since for 
scalar perturbations the flow is irrotational. Note that this definition is 
slightly different from the one used in Ref.~\cite{ML2}. The conservation of the 
energy--momentum tensors for each component provides the equations
\begin{align}
& \delta\ddot{\varphi}+3H\delta\dot{\varphi}
  +\frac{k^2}{a^2}\dph+\frac{\de^2 V}{\de \varphi^2}\dph
  \,=\, 4\dot{\varphi}\dot{\Phi}-2\frac{\de V}{\de \varphi}\Phi \comma
\label{e:EL}\\
& \dot{\delf}-3\gf\dot{\Phi}
  \,=\, \gf\frac{k^2}{a^2}\Vf \comma
\label{e:cont}\\
& \dot{\mathcal{V}}_\mathrm{f}
  \,=\, 3H\wf\Vf-\frac{\wf}{\gf}\delf-\Phi
\label{e:Euler}
\end{align}
where $\delf\equiv\drf/\rhf$. We also define $\delph\equiv\drph/\rhph$ and 
$\delpi\equiv\dpph/\rhph$. These equations are the perturbed Euler--Lagrange 
equation for the scalar field and the continuity and Euler equations for 
the fluid.

As we will see later, it is useful to introduce the comoving density 
perturbation for each component \cite{Bard}
\begin{equation}
\epsilon_\mathrm{x} \equiv \delta\rho_\mathrm{x}-3H\delta q_\mathrm{x} \comma
\end{equation}
which is a gauge-invariant quantity. We also introduce the gauge-invariant 
entropy perturbation variables \cite{KODA,WMLL,MWU}, namely the relative entropy 
perturbation between the fluid and the scalar field
\begin{equation}
\label{Sdef}
S \equiv
  \frac{3H\gf\gph\of}{\gamma}
  \(\frac{\drph}{\dot{\rhph}}-\frac{\drf}{\dot{\rhf}}\)
  = \of\frac{\gph\delf-\gf\delph}{\gamma} \comma
\end{equation}
and the intrinsic entropy perturbation of the scalar field
\begin{equation}
\label{Gdef}
\Gamma \equiv 
  \frac{3H\gph\csph}{1-\csph}
  \(\frac{\drph}{\dot{\rhph}}-\frac{\dpph}{\dot{\pph}}\)
  = \frac{\delpi-\csph\delph}{1-\csph} \comma
\end{equation}
where $\delpi\equiv\dpph/\rhph$. The normalizations have been chosen in order to 
simplify some later expressions. Note that $\Gamma$ is well defined even if 
$\csph\simeq1$ since, as can easily be shown, we have 
$\Gamma=\epsilon_\varphi/\rhph$. By definition the perfect fluid does not have 
an intrinsic entropy perturbation. Adiabaticity is defined by the condition 
$S=\Gamma=0$, since in this case it is possible to define a slicing for which 
all matter component perturbations vanish.

\subsection{Degrees of freedom}
\label{dof}

The system of Eqs.~(\ref{e:ein1}), (\ref{e:ein2}), (\ref{e:ein3}), (\ref{e:EL}), 
(\ref{e:cont}) and (\ref{e:Euler}) describes the evolution of $4$ variables, 
namely $\Phi$, $\rhf$, $\Vf$ and $\dph$. Equations (\ref{e:ein2}) and 
(\ref{e:EL}) are second order and if we introduce two new variables for 
$\dot{\Phi}$ and $\delta\dot{\varphi}$ (and therefore two new equations) we end 
up with $6$ variables describing the perturbations, 6 first-order dynamical 
equations and two constraint equations (Eqs.~(\ref{e:ein1}) and (\ref{e:ein3})). 
The two constraint equations reduce the number of degrees of freedom to $4$ and 
as a result two dynamical equations must be redundant. This comes from the fact 
that the conservation of the total energy--momentum tensor is a consequence of 
the Einstein equations and therefore the conservation equations for one matter 
component implies the ones for the other. As a result, it is possible to write 
this system as four differential equations for the four dynamical degrees of 
freedom which completely describe the perturbations, e.g. $\Phi$, $\delf$, 
$\delph$ and $\delpi$ which are the basic variables studied in Ref.~\cite{ML2} 
in the large-scale limit $k/aH \ll 1$. A general solution to those equations 
permits both an isocurvature perturbation between the scalar field density 
contrast and the fluid density contrast, and an isocurvature perturbation 
intrinsic to the scalar field, i.e. between its density and pressure 
perturbations.

For our discussion, it is useful to combine Eqs.~(\ref{e:ein1}) and 
(\ref{e:ein3}) and find the constraint equation
\begin{equation}
\label{constreq}
\frac{k^2}{a^2}\Phi = -4\pi G\epsilon_\mathrm{tot} \period
\end{equation}
Note that this is a gauge-invariant equation, though had we included an 
anisotropic stress then $\Phi$ must be replaced by the second metric potential 
$\Psi$ \cite{GWBM}. If the fluid is completely absent, so that we simply have a 
single scalar field, the constraint equation~\refeq{constreq} reduces to
\begin{equation}
\label{const_sf}
\frac{k^2}{a^2H^2}\Phi
= -\frac{4\pi G}{H^2} 
\(\dot{\varphi}\delta\dot{\varphi}-\dot{\varphi}^2\Phi-\ddot{\varphi}\dph\)
= - \frac32\Gamma \period
\end{equation}
The system is completely described by two dynamical degrees of freedom and this 
equation implies that one of the scalar field degrees of freedom is removed, 
e.g. $\delta\dot{\varphi}$. The right-hand side of \refeq{const_sf} is simply 
proportional to the intrinsic entropy perturbation of the scalar field $\Gamma$, 
hence in the large-scale limit this is forced to vanish if $\Phi$ is to remain 
small. This is a known result already shown in Refs.~\cite{BTKM,GWBM}.

By contrast, once a fluid is added we have
\begin{equation}
\label{constraint}
\frac{k^2}{a^2H^2}\Phi = -\frac32\[\oph\Gamma+\of\(\delf-3H\gf\Vf\)\] \period
\end{equation}
This equation shows that the fluid comoving density perturbation can compensate 
the scalar field intrinsic entropy perturbation, and, as a result, in the 
presence of a fluid it is possible to have a non-vanishing scalar field 
intrinsic entropy perturbation even on large scales. Note that the presence of 
the fluid changes the structure of the equations even if it is a sub-dominant 
component of the total energy density. This is because the fluid creates a new 
set of hypersurfaces, those on which its density is uniform, which need not 
align with hypersurfaces of uniform scalar field density.

Since we are interested in studying the evolution of isocurvature and adiabatic 
modes, we find useful to use the gauge-invariant comoving curvature 
perturbation \cite{MFB}
\begin{equation}
\label{R}
\R \equiv \frac{2(H\Phi+\dot{\Phi})}{3\gamma H}+\Phi \period
\end{equation}
The equation of motion for $\R$ is given by \cite{MFB} 
\begin{equation}
\dot{\R} = 
\frac{2}{3H\gamma}\[-\cs\frac{k^2}{a^2}\Phi+4\pi G\dpnad\] \comma
\label{rdot}
\end{equation}
where  $\dpnad\equiv\dptot-\cs\drtot$ is the non-adiabatic pressure 
perturbation. Note that even on large scales $\R$ can evolve due to the presence 
of a non-vanishing non-adiabatic pressure perturbation, as recently stressed in 
different works \cite{WMLL,MWU}. The non-adiabatic pressure perturbation 
depends on the intrinsic and relative entropy perturbations \cite{KODA,MWU}, and 
in our case we find
\begin{equation}
\frac{\dpnad}{\rhtot} = \oph\[(\wf-\csph)S+(1-\csph)\Gamma\] \period
\end{equation}
{}From now on we find it convenient to describe the system in terms of the 
gauge-invariant variables $\Phi$, $\R$, $S$ and $\Gamma$, rather than the set of 
variables $\Phi$, $\delf$, $\delph$, and $\delpi$. Note that such a change of 
variables is completely determined by Eqs.~(\ref{Sdef}), (\ref{Gdef}) and the 
expression
\begin{equation}
\R = \Phi
-\frac{1}{3 \gamma}\left( \Omega_{\varphi}\delta _{\varphi}+
\Omega_{\rm f}\delta _{\rm f} \right)-\frac{2}{9 \gamma}\frac{k^2}{a^2 H^2}\Phi 
\comma
\end{equation}
obtained from Eqs.~(\ref{e:ein1}) and (\ref{R}). In the next section we find 
a first-order system of dynamical equations expressed in these variables.

\section{Matrix formulation}

\subsection{Evolution equation}

In the long-wavelength limit ($k/aH\ll1$) Malquarti and Liddle \cite{ML2} were 
able to express the dynamical equations in a first-order matrix formulation, 
using $N\equiv\log(a/a_0)$ as a time variable. They took as basic variables 
$\Phi$, $\delf$, $\delph$, and $\delpi$. Here we show that our set of variables 
can bring the matrix into an even more efficient form. Moreover, we compute the 
general equations without the long-wavelength approximation.

We define the vector $\vv\equiv(\Phi,\R,S,\Gamma)^T$ and use a prime to 
denote a derivative with respect to $N$. Lengthy but straightforward algebra 
leads to the expression
\begin{equation}
\label{eqmat}
\vv'= \[\MM_0+\MM_1\frac{k^2}{a^2H^2}+\MM_2\frac{k^4}{a^4H^4}\]\times\vv
\end{equation}
where the only relevant matrix for the long-wavelength approximation ($k/aH=0$) 
is given by
\begin{widetext}
\begin{equation}
\label{mat}
\MM_0
=
\(\begin{array}{cccc}
 -(1+3\gamma/2) & 3\gamma/2 & 0 & 0 \\
 0 & 0 & \oph(\wf-\csph)/\gamma                   & \oph(1-\csph)/\gamma \\ 
 0 & 0 & 3(\wph-\wf)+3\gf\of(\wf-\csph)/\gamma & 3\gf\of(1-\csph)/\gamma \\
 0 & 0 & -3\gamma/2                               & 3(\wph-\gamma/2)
\end{array}\)
\comma
\end{equation}
\end{widetext}
and the two matrices incorporating spatial gradients are
\begin{equation}
\MM_1
=
\(\begin{array}{cccc}
 0             & 0     & 0    & 0    \\
 -2\cs/3\gamma & 0     & 0    & 0    \\ 
 0             & 0     & 1/3  & 1/3  \\
 0             & -\gph & -1/3 & -1/3
\end{array}\)
\comma
\end{equation}
and
\begin{equation}
\MM_2
=
\(\begin{array}{cccc}
 0 & 0 & 0 & 0 \\
 0 & 0 & 0 & 0 \\
 2\gph/9\gamma  & 0 & 0 & 0 \\
 -2\gph/9\gamma & 0 & 0 & 0
\end{array}\)
\period
\end{equation}

The different non-vanishing entries clearly show the couplings between adiabatic 
and relative/intrinsic entropy perturbations on large scales ($\MM_0$) and on 
small scales ($\MM_1$ and $\MM_2$). The first two lines of the matrices 
(dynamical equations for $\Phi$ and $\R$) are straightforward from 
Eqs.~(\ref{rdot}) and~(\ref{R}). The equation for the relative entropy, here 
expressed as
\begin{equation}
\label{sprime}
\begin{split}
S'& = \[3(\wph-\wf)+\frac{3\gf\of(\wf-\csph)}{\gamma}\]S \\
  & + \frac{3\gf\of(1-\csph)}{\gamma}\Gamma \\
  & + \frac{k^2}{a^2H^2}\[\frac13S+\frac13\Gamma\] 
    + \frac{k^4}{a^4H^4}\[\frac29\frac{\gph}{\gamma}\Phi\]
\comma
\end{split}
\end{equation}
has been obtained both in the context of multiple interacting fluids 
\cite{KODA,HN,MWU}, and in the framework of inflation when several interacting 
scalar fields are present \cite{GWBM}, \refeq{sprime} being a particular case. 
However, in general it is not possible to find a dynamical equation for the 
intrinsic entropy perturbation of a given component without knowing its 
underlying physics. In the case under study we are able to fully specify the 
evolution of the system through the equation for the intrinsic entropy 
perturbation of the scalar field as
\begin{equation}
\label{gamprime}
\begin{split}
\Gamma'& = -\frac32\gamma S+3\(\wph-\frac{\gamma}{2}\)\Gamma \\
       & +\frac{k^2}{a^2H^2}\[-\gph\R-\frac13S-\frac13\Gamma\] \\
       & +\frac{k^4}{a^4H^4}\[-\frac29\frac{\gph}{\gamma}\Phi\]
\period
\end{split}
\end{equation}
Eqs.~(\ref{sprime}) and~(\ref{gamprime}) show that, on large scales, the 
relative entropy perturbation and the intrinsic entropy perturbation of the 
scalar field are mutually sourced and evolve independently of the curvature 
perturbations. In particular, Eq.~(\ref{gamprime}) confirms the conclusions 
drawn from the constraint equation \refeq{constraint}, namely that in the 
presence of the fluid it is possible to have an intrinsic entropy perturbation 
relative to the scalar field even on large scales. When the fluid is very 
sub-dominant ($\of\simeq0$), we have $S\simeq0$ and therefore on large scales 
$\Gamma$ decays exponentially\footnote{Actually, $\Gamma$ remains constant in 
the special case $\wph=1$, but when the field is dominant this equation of state 
is usually not considered and anyway would rapidly evolve towards $\wph<1$.} 
with decay rate $-3+3\gph/2$, dynamically recovering the single scalar field 
case for which the intrinsic entropy perturbation vanishes (cf. 
Section~\ref{dof}).

When the fluid is completely absent, the matrices in \refeq{eqmat} reduce to 
$3\times3$ matrices, in the variables $\Phi$, $\R$ and $\Gamma$, but the 
constraint in \refeq{const_sf} allows to eliminate one more degree of 
freedom. For example, using \refeq{const_sf} one can find $2\times2$ matrices 
for $\R$ and $\Gamma$, or if one additionally goes to the large-scale limit, the 
constraint equation \refeq{const_sf} forces $\Gamma$ to vanish and gives $\MM_0$ 
as a $2\times2$ matrix for $\Phi$ and $\R$.

\subsection{Adiabatic condition}

The adiabatic condition requires that the relative entropy perturbation $S$ and 
the intrinsic entropy perturbation $\Gamma$ vanish. From our equations it is 
immediately clear that on large scales ($k/aH\ll1$, so that only $\MM_0$ need be 
considered)  if the perturbations are initially adiabatic then $S$ and $\Gamma$ 
remain zero for all times. In this case $\R$ is constant and $\Phi$ rapidly 
approaches its asymptotic value $\Phi=3\gamma \R/(2+3\gamma)$ (for constant or 
sufficiently slowly varying $\gamma$). This demonstrates that adiabaticity on 
large scales holds regardless of any time-dependence of the background variables 
such as $\wph$ and $\csph$ (this was already pointed out in Ref.~\cite{ML2}). In 
fact, preservation of adiabaticity is implied by the separate universe approach 
to large-scale perturbations \cite{WMLL}. However adiabaticity will be broken 
once the perturbations move out of the large-scale regime, with the matrices 
$\MM_1$ and $\MM_2$ sourcing $S$ and $\Gamma$ through the curvature 
perturbations $\R$ and $\Phi$. In particular, note that this is also true for 
the single scalar field case, as is evident from looking at \refeq{const_sf}. On 
the other hand, as we will see, if an isocurvature perturbation is initially 
present it can be wiped out on large scales by the scalar field dynamics.

Aspects of these results have appeared in previous works \cite{PB,AF,SA,DMSW}, 
but without noting that adiabaticity is always preserved on large scales. Our 
set of variables makes unambiguously clear the fact that the adiabatic condition 
is not an instantaneous notion on large scales, and holds independently of the 
evolution of the background.

\section{Application to the quintessence scenario}

\subsection{Analytical description}
\label{an_des}

In this section, we discuss the large-scale evolution of perturbations in 
quintessence scenarios. As described in the Appendix, before the quintessence 
field 
starts dominating the evolution of the universe its dynamics can feature up to 
four different regimes during which the coefficients of the matrix $\MM_0$ are 
constant. These are summarized in \reftab{val_reg}. Note that, as compared to 
Refs.~\cite{BMR} and \cite{ML2}, we altered the names of two regimes (potential 
I and II) in order to make our explanations clearer. Now, following 
Ref.~\cite{ML2}, for each regime it is possible to perform an eigenvector 
decomposition of the matrix $\MM_0$ and therefore compute analytically the 
large-scale evolution of the perturbations (i.e.~during each regime $\vv$ can 
be written as a sum of four terms proportional to $\vv_i\exp(n_iN)=\vv_ia^{n_i}$ 
for $i=1$ to $4$ respectively). However, the matching conditions between the 
different regimes are not obvious as $S$ and $\Gamma$ contain non-trivial 
functions of the background. In that respect, the formulation in Ref.~\cite{ML2} 
is more appropriate when following the modes over different regimes, since to a 
first approximation $\Phi$, $\delf$, $\delph$, and $\delpi$ can be taken as 
conserved through the transitions between regimes.

\begin{table}[t]
\begin{tabular}{l|ccc}
\hline
\hline
                       &
\makebox[1cm]{$\oph$}  & 
\makebox[1cm]{$\gph$}  &
\makebox[1cm]{$\csph$} \\
\hline
Kinetic               & 0      & 2      & 1         \\
Potential I           & 0      & 0      & 1         \\
Potential II          & 0      & 0      & $-2-\wf$  \\
Usual Tracker         & 0      & $\gph$ & $\wph$    \\
Perfect Tracker       & $\oph$ & $\gf$  & $\wph$    \\
\hline
\hline
\end{tabular}
\caption{Values of the three parameters $\oph$, $\gph$, and $\csph$ during the 
four different possible regimes of a quintessence scenario (a particular 
scenario would feature only one type of tracker regime) until the field 
starts dominating.}
\label{val_reg}
\end{table}

First of all, it is easy to find that for any regime $\MM_0$ possesses two 
eigenvectors
\begin{equation}
\begin{array}{lcl}
\vv_1 = (3\gamma,2+3\gamma,0,0) & \quad & n_1=0 \comma \\
\vv_2 = (1,0,0,0)               & \quad & n_2=-1-3\gamma/2 \comma
\end{array}
\end{equation}
where $n_\mathrm{x}$ is the eigenvalue of $\vv_\mathrm{x}$. These two vectors 
correspond to the two well-known adiabatic modes, the first one being constant, 
and the second one rapidly decaying.

Now, it is possible to find the two remaining entropy modes for a general case, 
but they cannot be expressed in a simple form. Nevertheless, it is 
straightforward and more clear to perform an eigenmode decomposition by 
considering each regime separately. The modes are given for each regime in 
\reftab{modes}. For simplicity, we do not display complicated expressions; for a 
detailed analysis the reader should refer to Ref.~\cite{ML2}. Most of these 
results have already been discussed in that paper, but here we would like to 
comment further in the light of our new set of variables and new findings.

\begin{table}[t]
\begin{tabular}{lcl}
\hline
\hline
&&\\
\multicolumn{3}{l}{Kinetic regime}\\
$\vv_3=(0,0,0,1)$        && $n_3=3-3\gf/2$ \\
$\vv_4=(0,0,2-\gf,\gf)$  && $n_4=0$ \\
&&\\

\multicolumn{3}{l}{Potential regime I}\\
$\vv_3=(0,0,0,1)$        && $n_3=-3-3\gf/2$ \\
$\vv_4=(0,0,2-\gf,\gf)$  && $n_4=-6$ \\
&&\\

\multicolumn{3}{l}{Potential regime II}\\
$\vv_3=(0,0,-2-\gf,\gf)$ && $n_3=0$ \\
$\vv_4=(0,0,-2,1)$       && $n_4=-3+3\gf/2$ \\
&&\\

\multicolumn{3}{l}{Tracker regime ($\gph\leq\gf$)}\\
$\vv_3=\ldots$           && $\Re(n_3)<0$ \\ 
$\vv_4=\ldots$           && $\Re(n_4)<0$ \\
&&\\

\hline
\hline
\end{tabular}
\caption{Eigenvectors and corresponding eigenvalues of the matrix $\MM_0$ of 
\refeq{eqmat} according to the different regimes.}
\label{modes}
\end{table}

As is well known, we see that entropy perturbations decay during the tracker 
regime \cite{AF,ML2}. This is due to the scaling and attractor properties of 
that regime. More striking is the fast-growing mode during the kinetic regime 
and the constant mode during the second potential regime. This contradicts the 
claim by Brax et al.~\cite{BMR} that the final value of the quintessence 
perturbations is insensitive to the initial conditions. The difference comes 
from the fact that these authors considered the basic variable $\delta\varphi$ 
and its time derivative. They found that there are two decaying modes for every 
regime. However, this does not mean that observationally-relevant variables 
(such as $\delph$) are decaying, since one has to take into account the 
evolution 
of background quantities (such as $\rho_\varphi$) as well. Our set of variables 
is therefore more appropriate.

Since in general we expect that there could be an initial relative entropy 
perturbation $S$ (for example in the case of a quintessence field present during 
inflation \cite{ML1}) and since $S$ sources $\Gamma$, we can expect a non-zero 
intrinsic entropy on large scales which would then evolve according to our set 
of equations Eqs.~(\ref{sprime}) and (\ref{gamprime}). Now, using the results 
given in the Appendix, Eqs.~(\ref{eqDNk}) and (\ref{eqDNpI}), we can show that 
the growth of $\Gamma$ during the kinetic regime --- $\exp(\DNk 
n_{3\mathrm{(k)}})=\exp(3\gf\DNp/2)$ --- is exactly compensated by its 
subsequent decay during the potential regime I --- $\exp(\DNpI 
n_{3\mathrm{(pI)}})=\exp(-3\gf\DNp/2)$. As a result, after the three 
regimes preceding the tracker regime, entropy modes are neither enhanced nor 
suppressed. However, as shown in the Appendix, according to the initial 
conditions, the potential energy of the field can undergo a very large drop 
before it reaches a constant value and the kinetic regime starts. In general, 
this transition phase could last a non-negligible number of $e$-foldings and 
would feature the same eigenmodes as the kinetic regime, in particular the same 
growing mode. Since this first phase of growth would not be compensated by the 
decay during the first potential regime, there remains the possibility that 
entropy modes may be enhanced at the beginning of the tracker regime. Now, if 
the tracker regime is not long enough to erase completely these entropy 
perturbations by the time the field becomes non-negligible, we may be able to 
see an imprint of these initial perturbations in observations \cite{AF,ML2}. As 
a result, we can see that in a quintessence scenario the initial conditions, as 
well as the history of the evolution of the background, are relevant when 
considering the late-time value of the perturbations.

In this respect many different assumptions can be made. In Ref.~\cite{KS}, 
Kneller and Strigari assumed equipartition as an initial value for the field, 
and in most cases this led to a field dynamics featuring a very long kinetic 
regime followed by the potential regimes and no tracker regime. In 
Ref.~\cite{Mac}, de la Macorra studied an actual physical model in which the 
quintessence field is a dark condensate which arises after a phase transition. 
Its evolution starts in the kinetic regime and again does not feature a tracker 
regime. In both scenarios entropy perturbations would still be present today. 
Alternatively, Malquarti and Liddle \cite{ML1} studied the evolution of a 
quintessence model during inflation in order to investigate the initial 
conditions of the quintessence field at the beginning of the radiation-dominated 
era. They found that typically the tracker starts at low redshift after a long 
period of potential regime II, but again, as a result, entropy perturbations 
generated during inflation could still lead to observable consequences. Finally, 
note that in general the initial entropy perturbations do not need to be of the 
order $\Phi$ and may be much larger.

\subsection{Numerical examples}
\label{num_ex}

In order to illustrate our results, we carried out simulations for two
quintessence models in a realistic universe (with radiation, dark matter and 
dark energy).

\begin{figure}[t]
\includegraphics[scale=0.34,angle=-90]{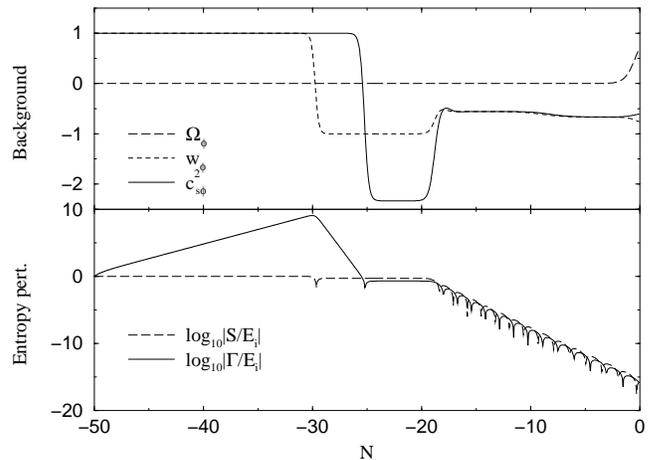}
\caption{Evolution in a realistic universe of background quantities (top) and 
entropy perturbation variables (bottom) of a quintessence field undergoing the 
four different possible regimes before its domination. We use an inverse 
power-law potential $V(\varphi)=V_0(\varphi/\mpl)^{-\alpha}$ (see 
Section~\ref{num_ex} for parameters). Note the transition between radiation 
domination and matter domination at $N\simeq-9$.}
\label{fig_entropy1}
\end{figure}

The first example is an inverse power-law model~\cite{RP}
$V(\varphi)=V_0(\varphi/\mpl)^{-\alpha}$, with $\alpha=1$ and 
$V_0=10^{-123}\,\mpl^4$, starting with the initial conditions at $N=-50$ given 
by $\rhph=10^{-20}\rhf$, $T=V$ and $S=-\Gamma=E_\mathrm{i}$. The results are 
shown in \reffig{fig_entropy1} where we display the evolution of some background 
variables and of the two entropy variables $S$ and $\Gamma$. After a very sharp 
transition towards the kinetic regime, the field undergoes the four regimes 
described in \reftab{val_reg} before starting dominating. First, we can clearly 
identify these regimes and see the growing, constant and decaying behavior of 
$S$ and $\Gamma$ corresponding to the modes displayed in 
\reftab{modes}.\footnote{Note that contrary to what the modes displayed in 
\reftab{modes} suggest, $S$ is constant during the potential regime I. This is 
because $\Gamma$ is many orders of magnitude larger than $S$, and $1-\csph$ is 
not exactly $0$.} We note the oscillations during the tracker regime and the 
non-trivial evolution through the transitions (yet keeping the same order of 
magnitude). We also observe that in this particular case the transition phase 
towards the kinetic regime is too short to enhance the entropy mode 
significantly, leading to the result that at the beginning of the tracker regime 
$S$ and $\Gamma$ have about the same amplitude as at the initial stage. In other 
words they are neither enhanced nor suppressed by the dynamics of the field, 
until tracking begins.

\begin{figure}[t]
\includegraphics[scale=0.34,angle=-90]{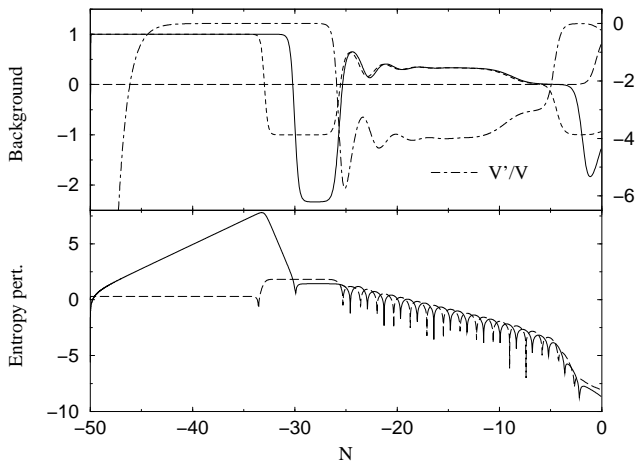}
\caption{As \reffig{fig_entropy1}, but now for a double-exponential potential 
$V(Q)=V_0[\exp(-\alpha\kappa\varphi)+\exp(-\beta\kappa\varphi)]$ (see 
Section~\ref{num_ex} for parameters). We also display $V'/V$ (to be read on the 
right-hand side of the graph).}
\label{fig_entropy2}
\end{figure}

The second example is a double-exponential model~\cite{BCN}
$V(Q)=V_0[\exp(-\alpha\kappa\varphi)+\exp(-\beta\kappa\varphi)]$, with 
$\alpha=1000$, $\beta=1$,  $V_0=10^{-122}\,\mpl^4$ and 
$\kappa=\sqrt{8\pi/3\mpl^2}$, starting with the initial conditions at $N=-50$ 
given by $\rhph=5\times10^{-3}\rhf$, $T=V$ and $S=E_\mathrm{i}$ and $\Gamma=0$. 
The results are shown in \reffig{fig_entropy2}. We display the same variables as 
in \reffig{fig_entropy1}. In addition, in order to observe the transition phase 
preceding the kinetic regime, we also display $V'/V$, where $V$ is the potential 
energy density of the field --- as described in the Appendix, the kinetic regime 
starts when $\abs{V'/V}$ drops under $\sim1$. First, we note that although 
$\Gamma=0$ initially, it evolves very quickly to be of the same order as $S$; 
as previously explained, this is due to the coupling between $S$ and $\Gamma$. 
Then, we observe the same behavior as for the first example, but this time we 
can see that the transition phase towards the kinetic regime lasts a few 
$e$-foldings. As a result, the entropy modes at the beginning of the tracker 
regime are nearly a factor $100$ larger than initially. This shows 
that, as discussed in Section~\ref{an_des}, entropy modes can be enhanced before 
the field reaches the tracker regime.

We have shown that, in general, one must take into account the initial 
conditions for the quintessence field and its perturbations in order to make any 
prediction.

\section{Discussion}

We have explored the nature of scalar perturbations for a universe filled with 
both a scalar field and a perfect fluid. We have introduced a useful set of 
variables and have provided a full analysis including spatial gradients. In 
particular, we have focused on the isocurvature perturbation modes and on the 
degrees of freedom which completely characterize the system. While for the case 
where only the scalar field is present its intrinsic entropy perturbation is 
forced to vanish at linear order for superhorizon scales, the presence of a 
fluid --- even if sub-dominant --- allows the possibility for such an intrinsic 
contribution to be present on large scales. However, in the case of a very 
sub-dominant fluid the intrinsic entropy of the scalar field decays, dynamically 
recovering the single scalar field situation.

We have recast the basic evolution equations in a rather simple matrix formalism 
in terms of the gauge-invariant variables for the adiabatic and isocurvature 
perturbations, taking into account the dynamics of the perturbations when a 
given wavelength re-enters the horizon. In particular, we have obtained an 
equation for the intrinsic entropy perturbation which shows that, on large 
scales, an initial adiabatic condition is indeed preserved, regardless of the 
evolution. Only when the perturbations approach the horizon are the adiabatic 
and entropy perturbations fully coupled together. 

Finally, we have applied our formalism to the quintessence scenario. In this 
case we have analyzed the large-scale evolution of the adiabatic and entropy 
perturbations in the different regimes which the quintessence scalar field 
dynamics may feature. As is well known, entropy perturbations are suppressed 
during the tracking regime, but, as already shown in Ref.~\cite{ML2}, during the 
kinetic regime one entropy mode undergoes an exponential growth. We have shown 
that it is then exactly compensated by an exponential decay during the first 
potential regime and then remains constant during the second potential regime. 
Therefore, after the three regimes preceding the tracker regime, entropy modes 
are neither enhanced nor suppressed. However, we discussed the remaining 
possibility of an enhancement during the short transition phase preceding the 
kinetic regime. We have studied two special cases numerically and have confirmed 
our analytical analysis. Moreover, we have observed that, in one of the cases, 
at the beginning of the tracker regime, entropy perturbations are larger than 
initially, and therefore we have concluded that entropy mode enhancement is 
possible. 

To summarize, we have shown that in general it is incorrect to assume that the 
observational imprint of quintessence perturbations will be independent of the 
initial conditions, because entropy perturbations can still be present when the 
quintessence energy density is no longer negligible. Note that this can happen 
as soon as the tracker starts, long before quintessence domination. In this 
case, entropy perturbations would feed curvature perturbations, but then 
would slowly decay to become negligible today.


\begin{acknowledgments}
N.B. was supported by \textsc{pparc}, P.S.C. by the Columbia University Academic 
Quality Fund, A.R.L. in part by the Leverhulme Trust and M.M. by the Fondation 
Barbour, the Fondation Wilsdorf, the Janggen-P\"{o}hn-Stiftung and the 
\textsc{ors} Awards Scheme. We thank Bruce Bassett for chairing an informal 
meeting at \textsc{cosmo03} on this topic, Fabio Finelli for discussions, and 
the referee for a report which motivated further study which significantly 
improved the paper. 
\end{acknowledgments}


\appendix

\section*{Appendix: Dynamical regimes of a quintessence field}

As described in Ref.~\cite{BMR}, a tracking quintessence field can feature up to 
four different dynamical regimes when in presence of a dominant fluid with 
constant equation of state. Here we clearly demonstrate the existence of these 
regimes and compute some relevant parameters. We will use the same notation and 
definitions as in the main body of the article.

\begin{figure}[t]
\includegraphics[scale=0.34,angle=-90]{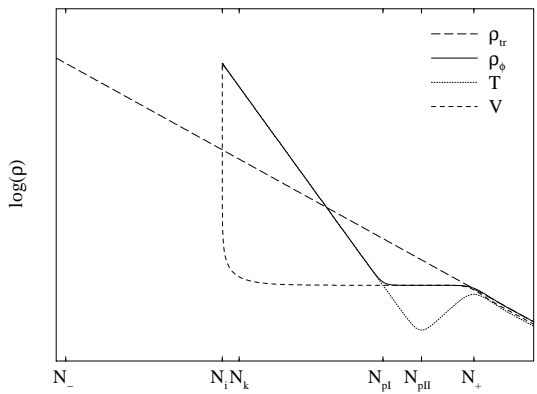}
\caption{Example of a quintessence field evolution featuring the four possible 
regimes before domination: kinetic, potential I, potential II and tracker. Note 
the transition phase towards the kinetic regime. The different parameters are 
described in the Appendix.}
\label{fig_regime}
\end{figure}

We assume that the quintessence field features a tracking solution $\rhot(N)$ 
--- note that it does not need to have a constant equation of state. This means 
that at each time (i.e.~for each value of $H$) there exists a stable field 
configuration for which its kinetic energy density $T\equiv\dot{\varphi}^2/2$ 
and its potential energy density $V$ are of the same order, and hence in 
\refeq{varphieq} the `friction' term due to the Hubble expansion and the slope 
of the potential balance each other. We will show that, according to the initial 
conditions, the scalar field can feature up to three different regimes before it 
reaches the tracker. We assume that the field is always subdominant and 
therefore does not influence the evolution of the universe, especially the 
evolution of $H$. As a result we have $H\propto\exp(-3\gf N/2)$. Using 
\refeq{varphieq} it is easy to see that
\begin{eqnarray}
\frac{T'}{T}&=&-\frac{V'}{T}-6\comma\label{eqT} \\
\frac{V'}{V}&=&\frac{\de V/\de\varphi}{V}\frac{\sqrt{2T}}{H}\period\label{eqV} 
\end{eqnarray}
In order to help the reader to follow the explanation, an example of a 
quintessence field evolution featuring the four possible regimes before 
domination is shown in \reffig{fig_regime}.

Let us start by looking at the initial condition $T\lesssim V\gg\rhot$ at time 
$\Ni$. The field, and hence the slope of the potential, has the same value as 
for the tracker configuration at an earlier time $\Nm<\Ni$, but because 
$H(\Ni)\ll H(\Nm)$ the friction term is actually negligible and the field 
fast-rolls down its potential ($V'\ll T'$) and its kinetic energy almost 
instantaneously dominates its energy density. At some time $\Nk$ shortly after 
$\Ni$ (at the latest when $T\sim\rhot$), the potential freezes at some value 
$\Vp\ll\rhot$ corresponding to the tracker configuration at a later time 
$\Np>\Nk$, and, since $H(N)\gg H(\Np)$ for $N<\Np$, it remains frozen until 
$\Np$. In this case, the evolution of $T$ can easily be computed analytically. 
We assume that at $\Np$ the tracker solution has an equation of state $\gtp$, 
express $\de V/\de\varphi$ as a function of $\Vp$ and $\gtp$, and for 
$\Nk<N<\Np$ we find 
\begin{equation}
\begin{split}
T(N) = & \;\; \frac{\gtp(2-\gtp)\Vp}{(\gf+2)^2}\times \\
      & \quad \left[Ce^{-3(N-\Np)} +e^{\frac32\gf(N-\Np)}\right]^2 \comma
\end{split}
\label{T_sol}
\end{equation}
where $C$ is a constant of integration depending on the initial conditions.

Let us explain this behavior. Starting from the time $\Nk$ the field is in the 
kinetic regime: $\rho_\varphi=T\propto\exp(-6N)$ and $V=\mathrm{constant}$. At 
some time $\NpI$ the field reaches the configuration $T\sim V$, and since the 
friction term is still extremely large ($H(\NpI)\gg H(\Np)$) the kinetic term 
keeps on decaying and the field enters the first potential regime: 
$\rho_\varphi=V=\mathrm{constant}$ and $T\propto\exp(-6N)$. At some time $\NpII$ 
the term $V'/T$ can balance the friction term and $T$ starts growing again. Here 
begins the second potential regime: $\rho_\varphi=V=\mathrm{const}$ and 
$T\propto\exp(3\gf N)$. Finally, at $\Np$ the field enters the tracker regime: 
$T\sim V\sim\rhot$. Note that the whole evolution described above goes through 
all the possible initial conditions.

Now we can compute a few parameters. First, using the solution for $T$ given in 
\refeq{T_sol} and noting that $\csph=-1-T'/3T$, it is straightforward to recover 
the values displayed in \reftab{val_reg}. In addition, let us define $\Delta 
N_\mathrm{r}$ as the number of $e$-foldings that the field spends in the regime 
`r' and also $\DNp\equiv\DNpI+\DNpII$. Using \refeq{eqV}, the fact that at $\Nk$ 
and $\Np$ we have $V'/V=-3\gtp\sim-1$ and the evolution for $T$ and $H$ 
described above, we find
\begin{eqnarray}
\DNk  &=& \frac{\gf}{2-\gf}\DNp \comma \label{eqDNk} \\
\DNpI &=& \frac{\gf}{2}\DNpII = \frac{\gf}{2+\gf}\DNp \period \label{eqDNpI}
\end{eqnarray}
These last two results are used in Section~\ref{an_des} to show that the growth 
of one of the entropy modes during the kinetic regime is exactly compensated by 
its subsequent decay during the first potential regime.


\end{document}